\documentclass[prb,superscriptaddress,twocolumn,floatfix]{revtex4}
\usepackage{graphicx}
\usepackage{amsmath,amssymb}
\usepackage{color}

\newfont{\fc}{cmssbx10 scaled 1000}

\begin{document}

\title{Direct observation of  dynamic charge stripes in La$_{2-x}$Sr$_{x}$NiO$_4$}

\author{S.~Anissimova}
\author{D.~Parshall}
\affiliation{Department of Physics, University of Colorado at Boulder, Boulder, Colorado 80309-0390, USA} 
\author{G. D. Gu}
\affiliation{Condensed Matter Physics \&\ Materials Science Department, Brookhaven National Laboratory, Upton, NY 11973-5000, USA} 
\author{K. Marty}
\author{M.~D.~Lumsden}
\author{Songxue~Chi}
\affiliation{Quantum Condensed Matter Division, Oak Ridge National Laboratory, Oak Ridge, Tennessee 37831, USA} 
\author{J.~A.~Fernandez-Baca}
\affiliation{Quantum Condensed Matter Division, Oak Ridge National Laboratory, Oak Ridge, Tennessee 37831, USA} 
\affiliation{Department of Physics and Astronomy, University of Tennessee, Knoxville, TN 37996-1200, USA}
\author{D.~L.~Abernathy}
\affiliation{Quantum Condensed Matter Division, Oak Ridge National Laboratory, Oak Ridge, Tennessee 37831, USA} 
\author{D.~Lamago}
\affiliation{CEA Saclay, Laboratoire L\'eon Brillouin, F-91191 Gif sur Yvette, France} 
\author{J. M. Tranquada}
\affiliation{Condensed Matter Physics \&\ Materials Science Department, Brookhaven National Laboratory, Upton, NY 11973-5000, USA} 
\author{D.~Reznik,$^{\ast}$}
\affiliation{Department of Physics, University of Colorado at Boulder, Boulder, Colorado 80309-0390, USA} 
\date{\today}


\maketitle

{\bf  The insulator-to-metal transition continues to be a challenging subject, especially when electronic correlations are strong. In layered compounds, such as La$_{2-x}$Sr$_x$NiO$_4$ and La$_{2-x}$Ba$_x$CuO$_4$, the doped charge carriers can segregate into periodically-spaced charge stripes separating narrow domains of antiferromagnetic order. Although there have been theoretical proposals of dynamically fluctuating stripes, direct spectroscopic evidence of charge-stripe fluctuations has been lacking. Here we report the detection of critical lattice fluctuations, driven by charge-stripe correlations, in La$_{2-x}$Sr$_x$NiO$_4$ using inelastic neutron scattering. This scattering is detected at large momentum transfers where the magnetic form factor suppresses the spin fluctuation signal.  The lattice fluctuations associated with the dynamic charge stripes are narrow in q and broad in energy. They are strongest near the charge stripe melting temperature.  Our results open the way towards the quantitative theory of dynamic stripes and for directly detecting dynamical charge stripes in other strongly-correlated systems, including high-temperature superconductors such as La$_{2-x}$Sr$_x$CuO$_4$.}

In a phase with a spontaneously broken symmetry, it is possible to have well-defined collective modes associated with the new order at temperatures far below the ordering temperature $T_c$.  For example, a system that develops magnetic order will typically exhibit transverse spin-wave excitations.  Above $T_c$, it is often possible to detect order-parameter fluctuations; if the interactions that drive the order are relatively isotropic, then the temperature range where these fluctuations are strong will be narrow.  In the case of lower-dimensional systems, however, it is possible to have a regime of order-parameter fluctuations that extend over a wide range of temperatures.  Examples include the layered, quasi-two-dimensional antiferromagnets K$_2$NiF$_4$ and La$_2$CuO$_4$, which exhibit dynamic antiferromagnetic spin correlations (with a temperature-dependent correlation length) to temperatures several times $T_c$.\cite{birg71,birg90,kast98}

On doping layered transition-metal-oxide antiferromagnets with holes, it is possible for a stripe phase to develop, with the holes segregating to charge stripes that form antiphase domain walls between antiferromagnetic strips.\cite{zaan01,kive03,vojt09,tran12c}  While the original proposals for the stripe phase focused on the ordered state,\cite{zaan89,mach89,kato90} it has been proposed that fluctuating order of this type (``dynamic charge stripes'') plays a key role in the physics of various interesting, strongly-correlated electronic materials.\cite{zaan01,cast95,kive98,frad10}  Here we report direct measurement of dynamic charge stripes by inelastic neutron scattering. 

\bigskip\noindent{\fc Results}\par

We have chosen to study the model system La$_{2-x}$Sr$_x$NiO$_4$ (LSNO), which is isostructural with the family of cuprates in which high-temperature superconductivity was first discovered.\cite{bedn86}  In particular, we focus on the dopant concentrations $x=0.25$ and $x=0.33$.\cite{lee97,du00,wu02}  The stripes develop within two-dimensional (2D) NiO$_2$ layers in which Ni atoms form a square lattice, with O atoms bridging the nearest neighbors.  The stripes form along the diagonal direction with respect to the Ni-O bonds.  While the average structure of both compositions is tetragonal, it is convenient to choose a unit cell containing two Ni atoms per layer, with in-plane lattice parameter $a\sim5.4$~\AA.  The 3D spin-order wave vector is then ${\bf q}_{\rm so}=(1\pm\delta,0,0)$,  while for charge order the wave vector is ${\bf q}_{\rm co}=(2\delta,0,1)$, in terms of reciprocal lattice units $(2\pi/a,2\pi/a,2\pi/c)$.  (The presence of stripe order should, through striction, cause the lattice parameters parallel and perpendicular to the stripes to become inequivalent.  This has not yet been detected in stripe-ordered LSNO, where the correlation length remains finite; however, stripes choose  a unique orientation in orthorhombic $R_{1.67}$Sr$_{0.33}$NiO$_4$ with $R=$ Pr, Nd.\cite{huck06})  The charge stripe period in real space equals $a/(2\delta)$.  Experiment\cite{yosh00} has shown that $\delta\approx x$.

Neutrons do not couple directly to charge.  The intensity at a charge-order superlattice peak is due to atomic displacements that help to screen the charge modulation.  In terms of the dynamical structure factor, $S({\bf Q},\omega)$, the static charge stripe order corresponds to an elastic peak, centered at $S({\bf G}\pm{\bf q}_{\rm co},0)$, where {\bf G} is a fundamental reciprocal lattice vector and the peak width is the inverse of the correlation length.  In contrast, the inelastic portion of $S({\bf Q},\omega)$ for {\bf Q} close to ${\bf G}\pm{\bf q}_{\rm co}$ is not strongly peaked at $\omega=0$; its {\bf Q}-space structure reflects the periodicity and correlation length, and its frequency dependence reflects the persistence time of the fluctuating order. 

\begin{figure}[t]
\centerline{\includegraphics[width=\columnwidth]{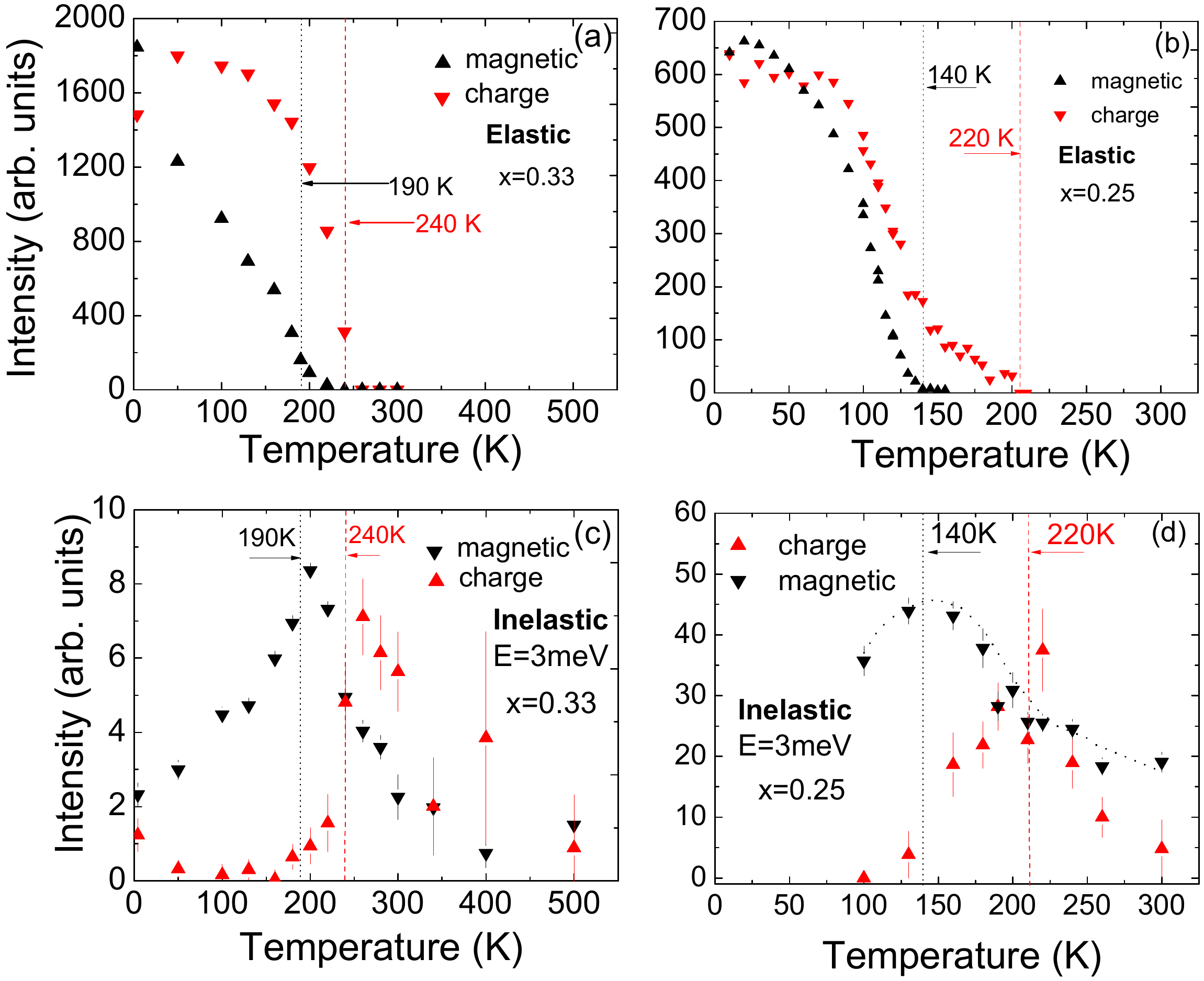}}
\caption{
{\fc Temperature dependence of static and dynamic stripe correlations in La$_{2-x}$Sr$_{x}$NiO$_4$.}  Spin (black) and charge (red) Bragg peak intensities for (a) $x=0.33$, (b) $x=0.25$.  Magnetic (black) and charge-stripe (red) contributions to inelastic scattering at $E=3$~meV for (c) $x=0.33$, (d) $x=0.25$.  Analysis behind these quantities is explained in the text.  In all panels, vertical lines indicate spin and charge ordering temperatures.
\label{fg:Bragg_Inel}}
\end{figure}

To distinguish the dynamic lattice response associated with charge stripe fluctuations from the spin fluctuations known to be present,\cite{tran97c,lee02,bour03} we rely on distinct signatures of the two types of scattering as a function of  both $|{\bf Q}|$ and $Q_z$; we will explain these distinctions in the following.  The charge and spin order parameters have distinct temperature dependences: Magnetic and charge Bragg peaks disappear at different temperatures, T$_{\rm so}$ and T$_{\rm co}$ respectively, as shown in Fig.~\ref{fg:Bragg_Inel}a,b. The charge and spin stripe fluctuations also have distinct temperature dependences, as summarized in Fig.~\ref{fg:Bragg_Inel}c,d.  For fluctuations at an excitation energy of 3~meV, the charge and spin responses each peak near the temperature where the associated order disappears, $T_{\rm co}$ and $T_{\rm so}$.  The charge fluctuations, in particular, are detected over a wide range of temperature above $T_{\rm co}$. 

\begin{figure}[t]
\centerline{\includegraphics[width=0.8\columnwidth]{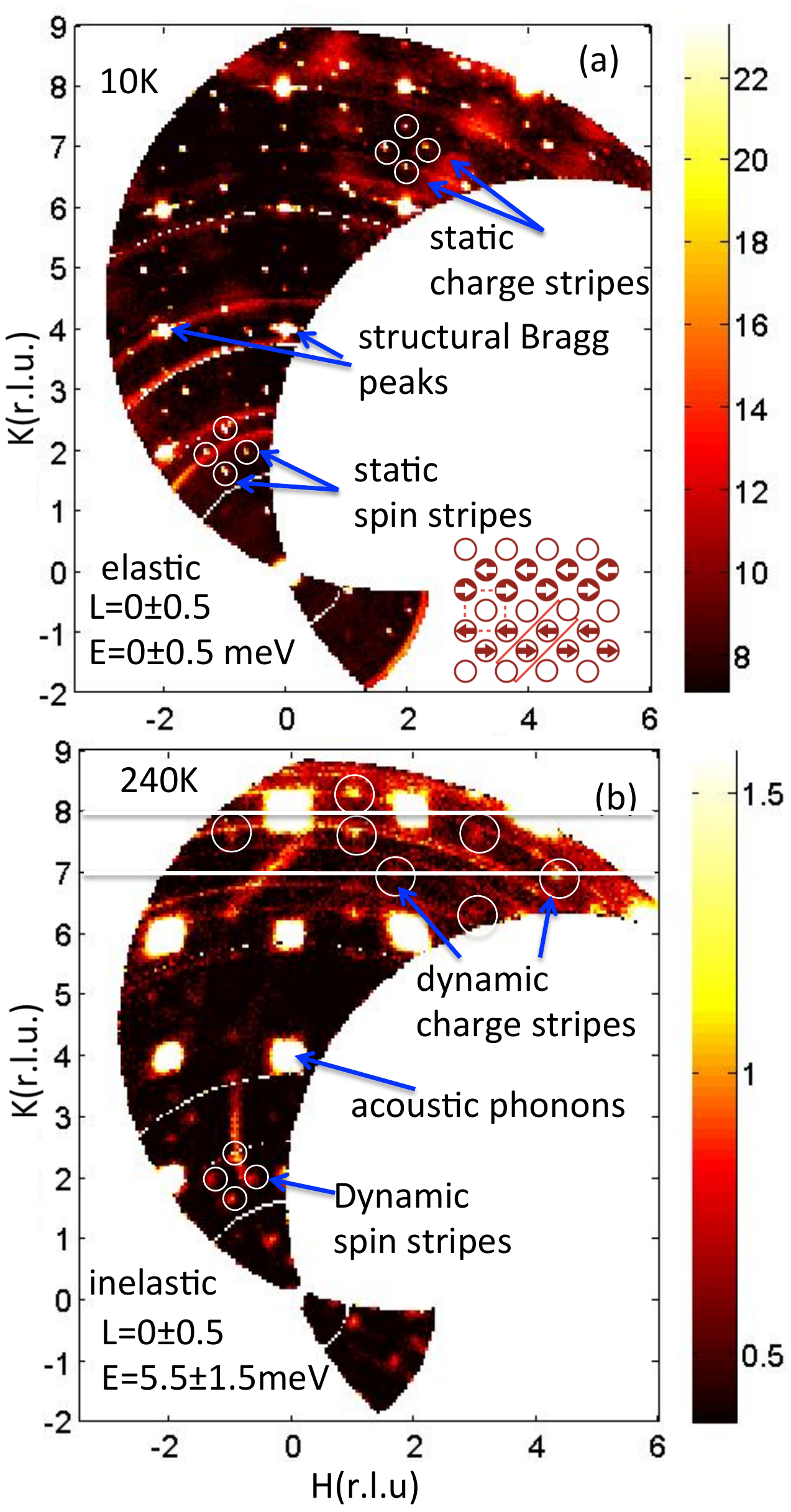}}
\caption{
{\fc Reciprocal-space maps for La$_{1.67}$Sr$_{0.33}$NiO$_4$.} {\fc a} Elastic scattering in the $(h,k,0)$ plane measured at 10~K.  Commensurate Bragg peaks are due to average lattice; quartets of incommensurate peaks are due to stripe order: high-$Q$ (low-$Q$) peaks dominated by charge-related (magnetic) scattering.  (Arcs are powder diffraction from aluminum sample holder.)  {\fc b} Inelastic scattering ($4 \le E\le 7$ meV), measured at $T_{\rm co}$.  Bright patches are acoustic phonons.   Weak peaks at large (small) $k$, denoted by circles, originate from dynamic charge (spin) stripes. Lines at the top of panel  indicate $k = 7$ and 8. (Streaks are contaminations from accidental Bragg scattering by sample.)  {\fc a inset} Sketch of stripe order, showing only Ni sites.  Open circle: Ni$^{2+}+$ hole; arrows indicate orientation of ordered moments, with change in shading indicating the $\pi$-shift in antiferromagnetic phase across each charge stripe.  Red dashed line: unit cell for indexing purposes; red solid line: unit cell for both spin and charge order.
\label{fg:map}}
\end{figure}

\noindent{\bf Neutron scattering study on an optimally-doped sample.}
We began our investigation with the $x=0.33$ sample, the optimally-doped composition corresponding to the highest magnetic and charge-ordering temperatures in the La$_{2-x}$Sr$_x$NiO$_4$ family.\cite{yosh00} For $x=\delta=\frac13$, the spin and charge modulations are characterized by the same size unit cell, as indicated in the inset of Fig.~\ref{fg:map}a; it follows that diffraction measurements of these modulations are characterized by the same wave vector, ${\bf q}_{\rm s}$.  The modulation wave vector for a 2D layer is then ${\bf q}_{\rm s} = (0,\frac23)$.  Because of the tetragonal symmetry of the average lattice, it is equally probable to have stripe domains rotated by $90^\circ$ and characterized by ${\bf q}_s'=(\frac23,0)$.  For simplicity, we will use ${\bf q}_{\rm s}$ to represent the wave vectors for both stripe domains.

Because of the coincidence of the charge and spin wave vectors in this case ($1-\delta=2\delta$ when $\delta=\frac13$), special care has to be taken to properly separate the scattering associated with charge fluctuations from the magnetic response.  The magnetic scattering intensity is proportional to the square of the magnetic form factor, which decreases with $Q$ and has fallen to 25\%\ of its maximum by $Q\sim 5a^*$.\cite{wang92} In contrast, inelastic scattering from lattice deformations increases with $Q$ (in fact, as $Q^2$ if the structure factor is ignored). So one way to isolate dynamic charge stripes from spin stripes is by measuring the former at large momentum transfers.

Figure~\ref{fg:map}a shows an intensity map for elastic neutron scattering measured at 10~K, which is well below the charge-ordering temperature,\cite{lee97} $T_{\rm co} = 240$~K.  It is plotted in the  $(H,K,0)$ plane of reciprocal space, with $L$ integrated from $-0.5$ to 0.5 and $E$ integrated from $-0.5$ to 0.5~meV.  The energy interval is smaller than the experimental resolution, so this cut probes only elastic ($E=0$) diffraction intensity. In terms of the $F4/mmm$ space group, the crystal structure of La$_{2-x}$Sr$_x$NiO$_4$ allows fundamental Bragg peaks with indices $H$, $K$, $L$ all even or all odd.  Strong fundamental Bragg peaks, associated with the average atomic structure, appear at ${\bf Q} = {\bf G} = (2m,2n,0)$, with $m$, $n=$ integers.  Spin-order peaks appear at ${\bf G}\pm{\bf q}_{\rm so}$.  Charge-order peaks of the type ${\bf G}\pm{\bf q}_{\rm co}$ will appear in the $L=\pm1$ planes, but not at $L=0$; however, there are also fundamental Bragg peaks at ${\bf G'}=(2m+1,2n+1,\pm1)$, so that charge order peaks at ${\bf G'}\pm{\bf q}_{\rm co}$ appear in the $L=0$ plane, overlapping spin-order peaks.  To summarize, peaks at $(2n\pm\frac23,0,0)$, or equivalently $(0,2n\pm\frac23,0)$ should be spin only (if the peak widths along $L$ are narrow\cite{lee97}), while all other superlattice peaks in the $L=0$ plane have mixed spin and charge contributions.

As discussed above, scattering from dynamic stripes should appear in the inelastic channel.  At 10~K, the low-energy inelastic scattering near the superlattice wave vectors falls off with increasing $Q$ in a manner consistent with the magnetic form factor, and it is independent of $L$.  These observations indicate fluctuations from the spin stripes,\cite{lee02,woo05}  with no dynamic spin correlations between the NiO$_2$ layers.  Charge-stripe fluctuations become apparent in the inelastic data ($4<E<7$~meV) shown in Fig.~\ref{fg:map}b, measured at $T=T_{\rm co}$, where the elastic charge-stripe peaks disappear.  Here, the strongest scattering comes from acoustic phonons centered on wave vectors ${\bf G}$.  One can also see weak scattering about some of the superlattice positions ${\bf G}'\pm{\bf q}_{\rm s}$.  The excitations seen at smaller $|{\bf Q}|$ values are due to spin fluctuations; note that the spin-only scattering at $(0,5\pm\frac13)$ is negligible due to the magnetic form factor.  

\begin{figure}[t]
\centerline{\includegraphics[width=\columnwidth]{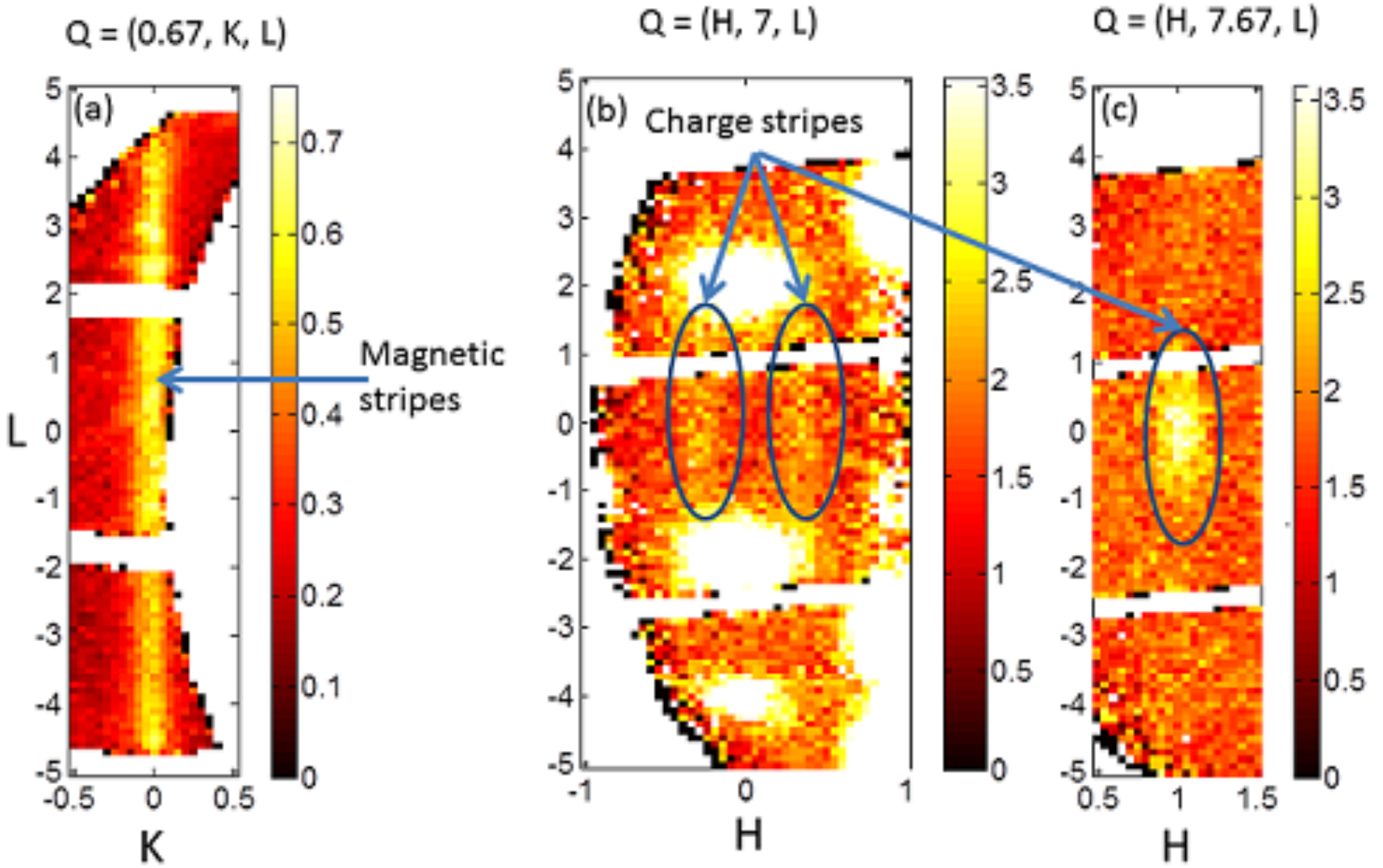}}
\caption{
{\fc Distinct $L$-dependence of spin and charge-stripe fluctuations.}   {\fc a} Magnetic scattering at ${\bf Q}=(0.67,0,L)$.   Charge stripe fluctuations at  {\fc b}: ${\bf Q} = (\pm 0.33, 7, L)$ and {\fc c}: ${\bf Q} = (1, 7.67,L)$, as indicated by ovals.  Gaps in the data are due to gaps between detector banks.  All measurements are at $T=240$~K and $E = 5.5 \pm 1.5$~meV. 
\label{fg:ldep} }
\end{figure}

As a further test, representative examples of the $L$-dependence of scattering from small-$Q$ and large-$Q$ positions are shown in Fig.~\ref{fg:ldep}.  The small-$Q$ example, Fig.~\ref{fg:ldep}a, shows no structure in $L$, consistent with the response from 2D spin fluctuations.  In contrast, the large-$Q$ examples exhibit a broad peak about $L=0$.  This is consistent with lattice distortions induced by charge modulation, as the structure factor involves atoms both within the NiO$_2$ planes and outside of them, such as the apical O atoms.  (We will give further examples below for the case of $x=0.25$.)  Thus, we conclude that the peaks at very large $|{\bf Q}|$, with resurgent intensity that is modulated in $L$, are associated with charge-stripe fluctuations.

\begin{figure}[b]
\centerline{\includegraphics[width=\columnwidth]{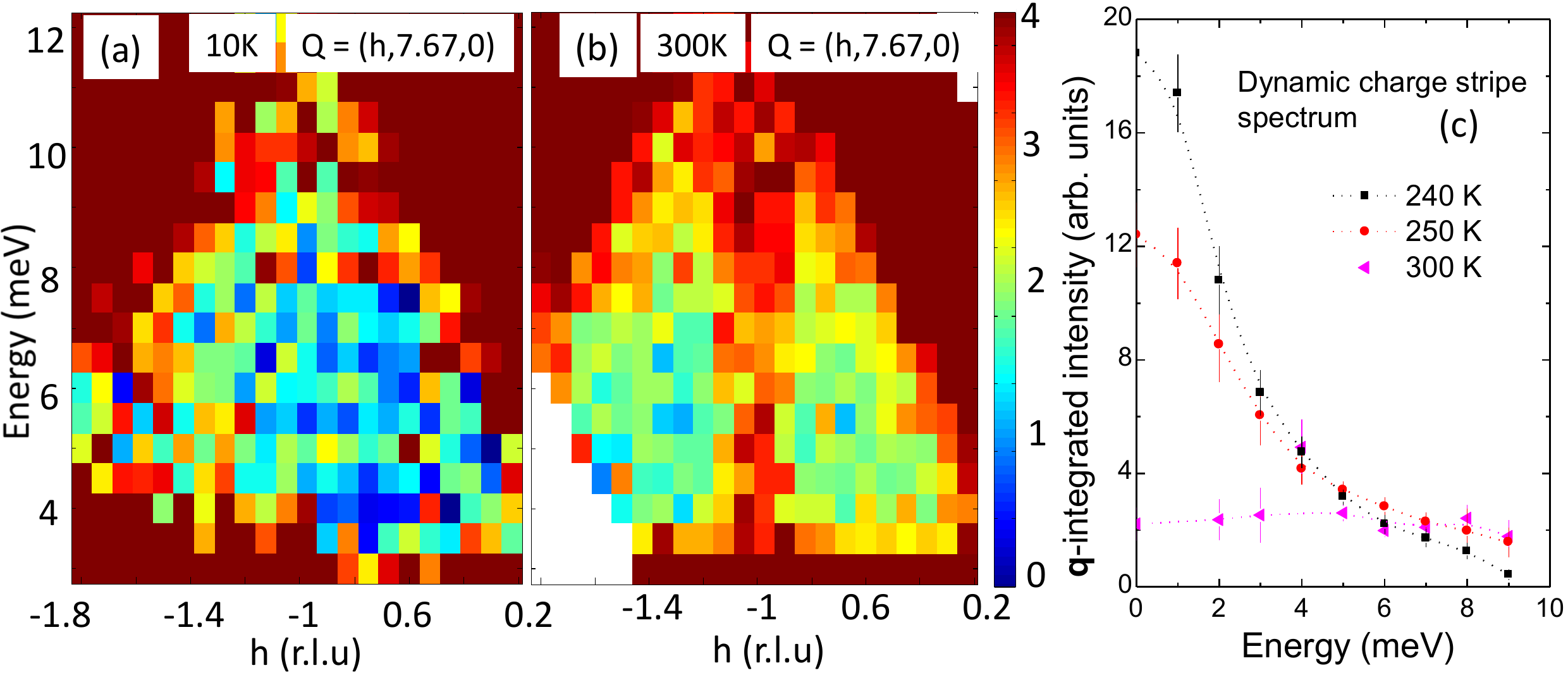}}
\caption{
{\fc Energy-dependence of charge-stripe fluctuations in La$_{1.67}$Sr$_{0.33}$NiO$_4$.} $E$ vs.\ $h$ cuts through ${\bf Q}_0=(-1,7.67,0)$ at {\fc a}: 10~K and {\fc b}: 300~K.  {\fc c} Integrated intensity of the charge-stripe fluctuations at ${\bf Q}_0$ plotted vs.\ energy for several temperatures.  Error bars represent statistical error.  Energy resolution is $\sim2$~meV half-width at half-maximum.
\label{fg:mapE}}
\end{figure}

Having determined the character of the high $Q$ excitations, we next consider their energy dependence.  Figure~\ref{fg:mapE}a and b compare measurements about ${\bf Q}=(-1,7)+{\bf q}_{\rm s}$ at temperatures of 10~K and 300~K, respectively. At this large {\bf Q} the magnetic form factor is negligibly small and all the signal originated from the atomic lattice fluctuations induced by dynamic charge stripes. Here the data are plotted as a function of $h$ and $E$.  At 300~K, well above $T_{\rm co}$ one can see excitations rising almost vertically at $h=-1$, while the same excitations are undetectable above the background at 10~K.  The temperature-dependence of excitations for $T_{\rm co}\le T\le300$~K is shown in Fig.~\ref{fg:mapE}c, where the intensity as a function of $E$  has been integrated over {\bf Q} about $(-1,7.67)$, and the background has been subtracted.  While the low-energy excitations become strongly enhanced on cooling towards $T_{\rm co}$, developing a quasi-elastic distribution, the excitations at 300~K clearly indicate a dynamical charge-stripe phase.

To obtain a better measure of the temperature dependence of the excitations, a second set of measurements was done on a different instrument using a lower incident energy.  This provided better energy resolution, but also limited the $Q$ range that could be reached.  As a consequence, it is necessary to correct these data for coexisting spin- and charge-stripe contributions, as we will explain.

The measurements were performed at one superlattice position dominated by spin scattering, ${\bf Q}_1=(0.33,1,0)$, and a second dominated by charge-stripe scattering, ${\bf Q}_2=(3.67,1,0)$.  The temperature dependence of the elastic scattering at these positions is shown in Fig.~\ref{fg:Bragg_Inel}a (labeled ``magnetic'' and ``charge'', respectively).  One can see that the ${\bf Q}_1$ signal disappears on warming through the spin-ordering temperature, $T_{\rm so}=190$~K, whereas the ${\bf Q}_2$ peak drops off at $T_{\rm co}$, consistent with previous studies.\cite{lee97,rami96} 

\begin{figure}[t]
\centerline{\includegraphics[width=\columnwidth]{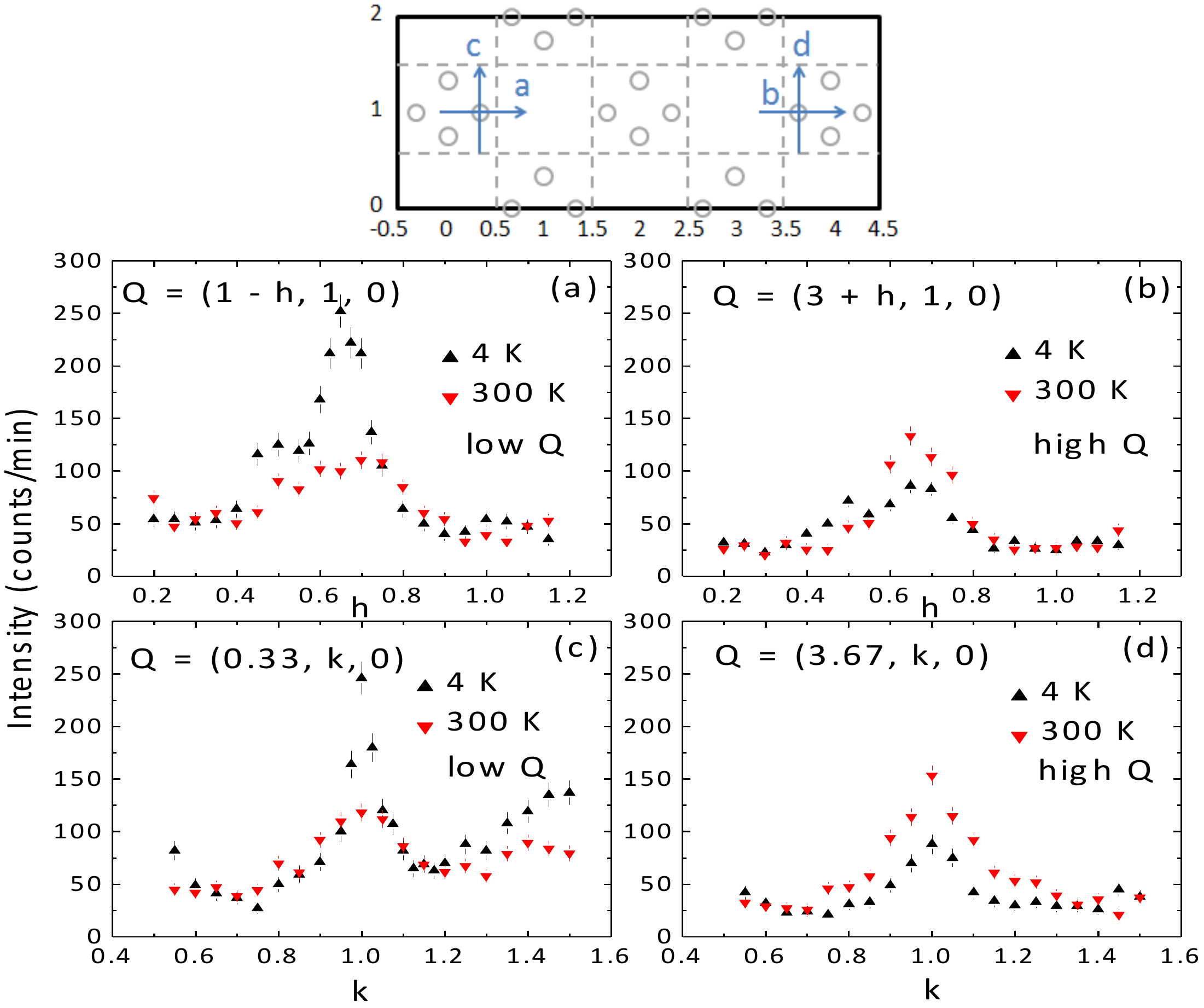}}
\caption{
{\fc Scans through spin and charge excitations.}   
{\fc a, c}  Scans through ${\bf Q}=(0.33,1,0)$; {\fc b, d} scans through ${\bf Q}=(3.67,1,0)$, all at $E=3$~meV.   Scan directions are indicated in the inset at top.  Measurements at $T=4$~K (black) and 300~K (red).  
\label{fg:peaks}  }
\end{figure}

The inelastic scattering at these positions was measured along orthogonal cuts in reciprocal space, as indicated in the top inset of Fig.~\ref{fg:peaks}.  Panels a and c show the scans at $E=3$~meV obtained at $T=4$~K  and 300~K along the orthogonal directions through ${\bf Q}_1$.  The signal here should be dominated by spin fluctuations, and the decrease in peak intensity at 300 K compared to 4 K is consistent with that expectation.  (Note that the peaks observed when the scans cross the dashed lines of the inset correspond to one-dimensional spin correlations within the charge stripes; see Boothroyd {\it et al.}\cite{boot03b}) Panels b and d show scans through ${\bf Q}_2$.  The scattering near ${\bf Q}_2$ should be dominated by charge-stripe correlations.  Indeed, we see a very different behavior from ${\bf Q}_1$: the inelastic intensity {\it increases} at 300~K compared to 4~K. 

\begin{figure}[t]
\centerline{\includegraphics[width=\columnwidth]{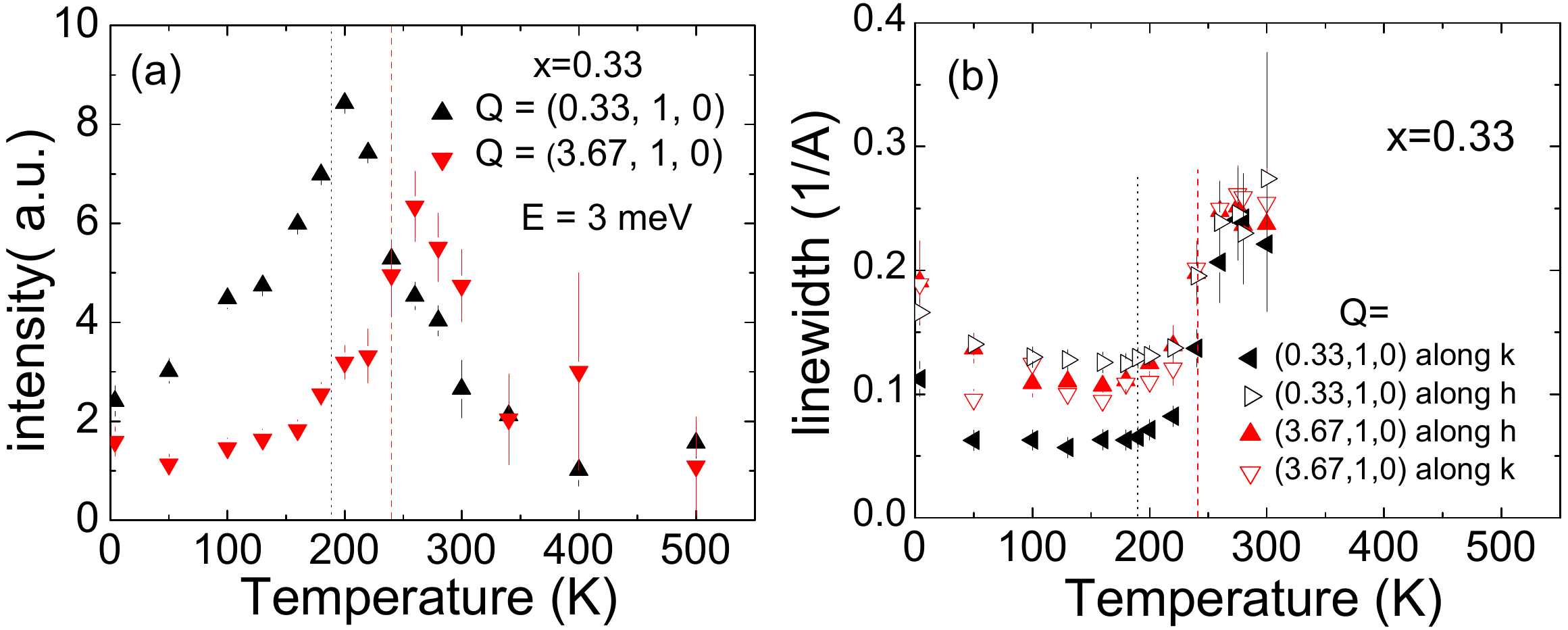}}
\caption{
{\fc Temperature dependence of spin and charge excitations.}   
{\fc a}  Inelastic Intensities at ${\bf Q}=(0.33,1,0)$ (black) and ${\bf Q}=(3.67,1,0)$ (red), for $E=3$~meV.  {\bf b} Corresponding line widths at ${\bf Q}=(0.33,1,0)$ (black) and ${\bf Q}=(3.67,1,0)$ (red).
\label{fg:intens}  }
\end{figure}

The full temperature dependence of the integrated intensities at $E=3$~meV is presented in Fig.~\ref{fg:intens}a.  The fitted peak widths are plotted in Fig.~\ref{fg:intens}b; above 300~K, where the peaks have become broad and the peak intensity becomes comparable to background, the widths were fixed (and hence are not shown) in order to obtain a reasonable integrated intensity.   Focusing on the intensities in Fig.~\ref{fg:intens}a, we see that the inelastic intensity at ${\bf Q}_1$ peaks just above $T_{\rm so}$, while the signal at ${\bf Q}_2$ peaks near $T_{\rm co}$.

\begin{figure}[b]
\centerline{\includegraphics[width=0.9\columnwidth]{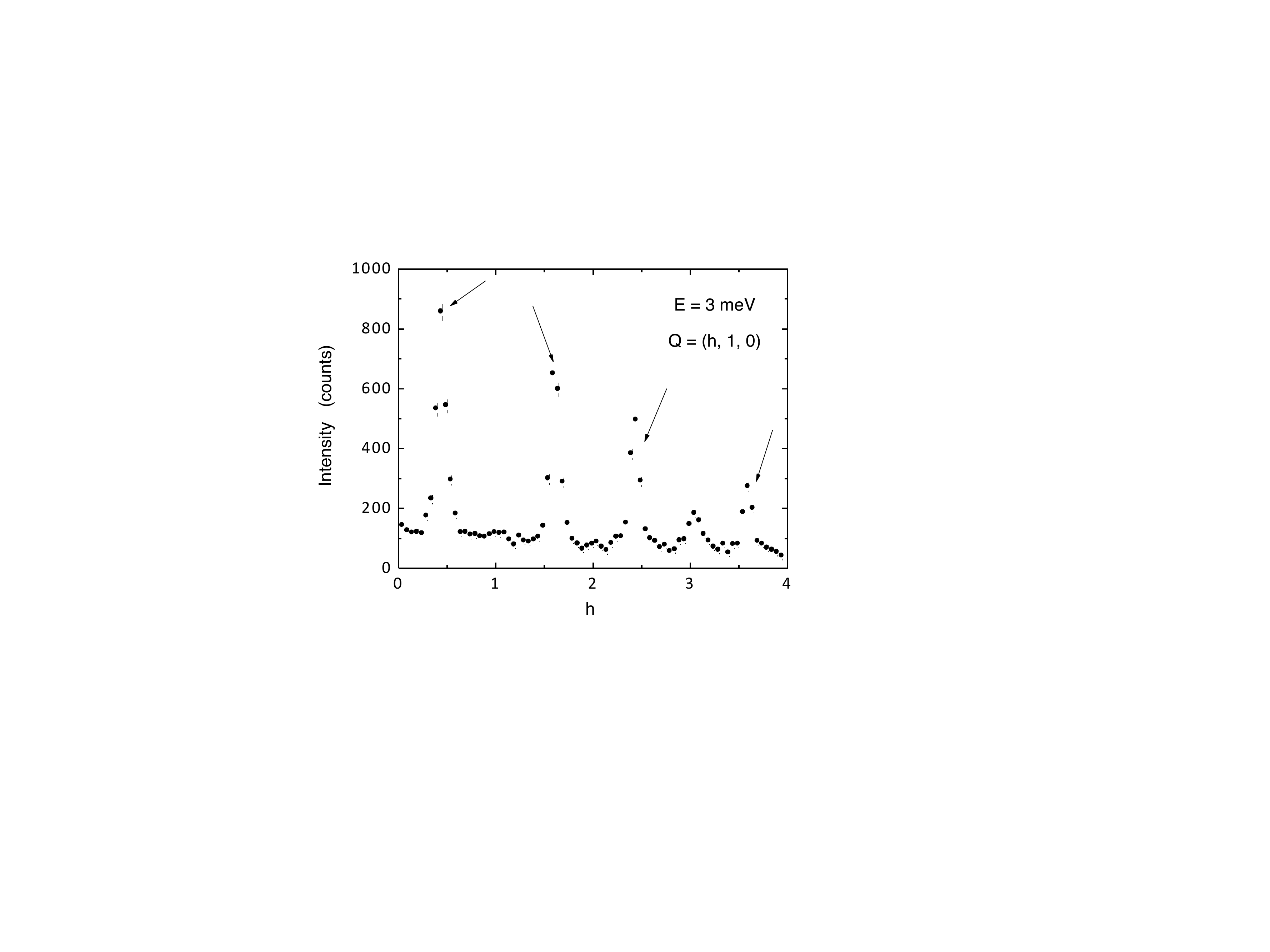}}
\caption{
{\fc Determination of the magnetic form factor.}   
Constant-energy scan at  $E = 3$~meV along ${\bf Q}=(h,1,0)$ for LSNO with $x=0.45$ at $T=160$~K. Arrows point to the magnetic peaks.  The peak at $Q=(3.05,1,0)$ originates from acoustic phonons associated with the (3,1,1) Bragg peak.
\label{fg:x45}  }
\end{figure}

To be more quantitative, we need to correct for mixed contributions.  Let $M$ and $C$ represent the magnetic and charge contributions respectively; then, we can write
\begin{eqnarray*}
 I({\bf Q}_1) & = & C/A + M,\\
 I({\bf Q}_2) & = & C + M/F^2,\\
\end{eqnarray*}
where $F^2$ accounts for the change in the magnetic form factor (as well as differences in geometrical factors associated with the scattering cross section for spin fluctuations).   To get an estimate of $F^2$, we make use of inelastic scattering data obtained on a crystal of LSNO with $x=0.45$, for which the magnetic scattering is well separated in {\bf Q} from the charge-stripe scattering.  Figure~\ref{fg:x45} shows a scan at $E=3$~meV along ${\bf Q}=(h,1,0)$.  From the peak intensities at $h=0.45$ and 3.45, which are close to our ${\bf Q}_1$ and ${\bf Q}_2$, we obtain $F^2=3.5\pm0.5$.  The parameter $A$ accounts for the difference in scattering strength for the atomic displacements.  We set $A=10$, which roughly corresponds to $Q_2^2/Q_1^2$; the precise value has little impact on our results as long as $A\gg 1$.  Inverting the pair of equations yields the results for $C$ and $M$ plotted in Fig.~\ref{fg:Bragg_Inel}c.

\begin{figure}[b]
\centerline{\includegraphics[width=\columnwidth]{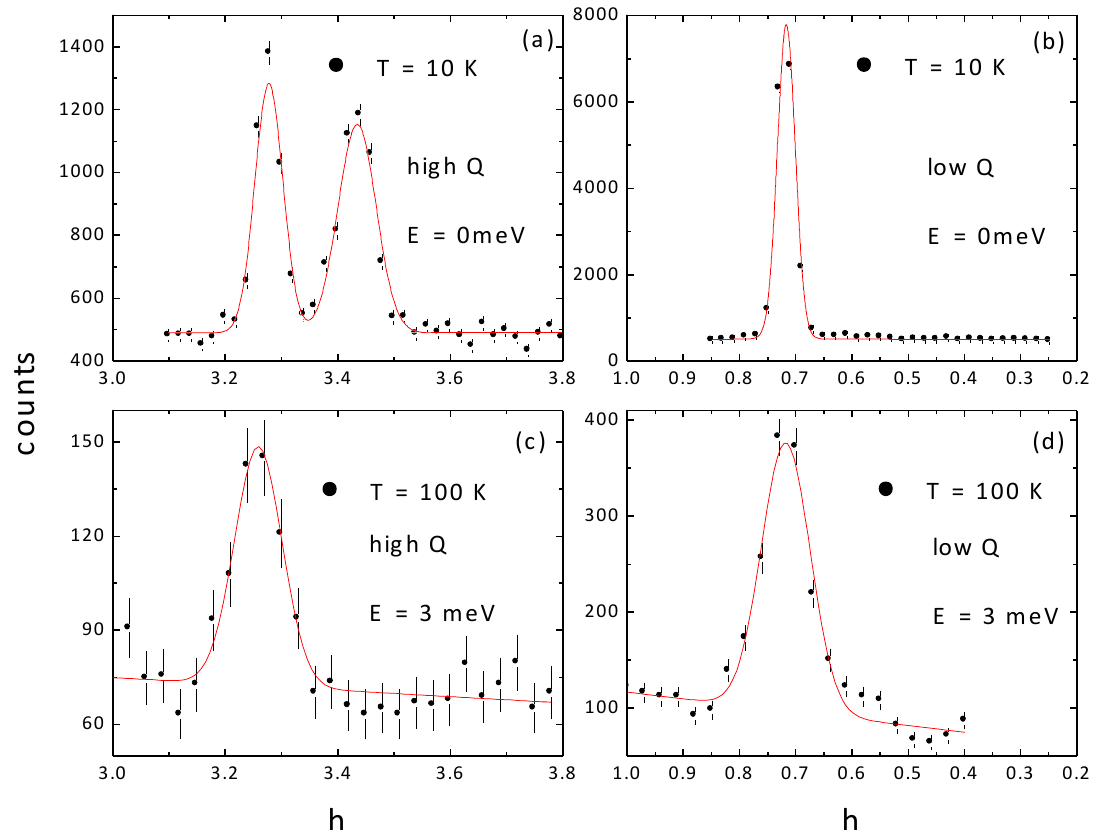}}
\caption{
{\fc Scans of spin and charge stripe order and fluctuations in La$_{1.75}$Sr$_{0.25}$NiO$_4$. }   
{\fc a, b} Elastic scans along $(h,0,1)$ measured $T=10$~K for the same fixed counting time.  In (a), one can see a magnetic peak at $h=3+\delta$ and a charge order peak at $h=4-2\delta$, while (b) shows only a magnetic peak at $h=1-\delta$.  {\fc c, d} Inelastic scans at $E=3$~meV along $(h,0,1)$ for the same fixed counting time, measured at $T=100$~K to enhance the inelastic intensity.  Signal is seen only at the wave vectors corresponding to spin order.  In each panel, the red line is a fit of one or more Gaussian peaks on top of a linear background. 
\label{fg:el1}  }
\end{figure}

\noindent{\bf Neutron scattering study of an underdoped sample.}
To confirm our identification of charge-stripe fluctuations, we have done related measurements on an underdoped crystal of LSNO with $x=0.25$.  In this case, the representative wave vectors studied are ${\bf Q}_1'=(1-\delta,0,1)$ and ${\bf Q}_2'=(4-2\delta,0,1)$.  From the elastic scans at 10~K shown in Fig.~\ref{fg:el1}b and a, respectively, we see that $\delta=0.28$ and that the charge-stripe peak at ${\bf Q}_2'$ has a magnetic neighbor at $(3+\delta,0,1)$.  The temperature dependences of the ${\bf Q}_1'$ and ${\bf Q}_2'$ peak intensities are plotted in Fig.~\ref{fg:Bragg_Inel}b as ``magnetic'' and ``charge'', respectively.  We see that $T_{\rm so}\approx 140$~K and $T_{\rm co}\approx220$~K.  Inelastic scans at $E=3$~meV for $T=100$~K,  shown in Fig.~\ref{fg:el1}d and c, indicate that below $T_{\rm so}$ only magnetic fluctuations are readily apparent.

\begin{figure}[t]
\centerline{\includegraphics[width=\columnwidth]{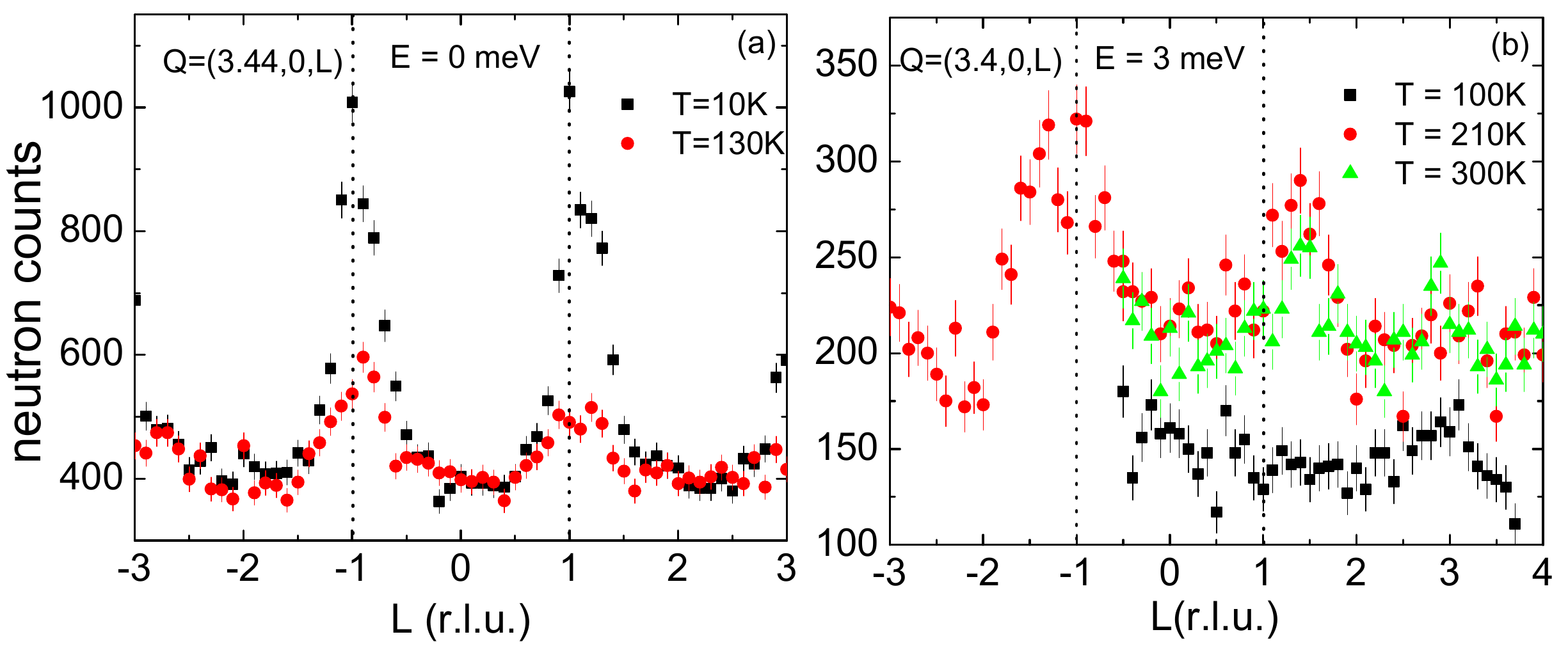}}
\caption{
{\fc $L$ dependence of charge-stripe scattering in La$_{1.75}$Sr$_{0.25}$NiO$_4$.} 
{\fc a} Elastic scattering along {\bf Q}=(3.44,0,L) at $T=10$~K (black) and 130~K (red).  {\fc b} Inelastic scattering ($E=3$~meV) along {\bf Q}=(3.4,0,L) at $T=100$~K (black), 210~K (red), and 300~K (green).  Dashed lines at $L=\pm1$ are guides to the eye.
\label{fg:Ldep25}  }
\end{figure}

The $L$-dependence of the charge-stripe scattering measured around ${\bf Q}_2'$ is presented in Fig.~\ref{fg:Ldep25}.  Panel a shows the elastic scattering at two temperatures below $T_{\rm so}$; the fact that the scattering peaks at $L=\pm1$ and $\pm3$ is controlled by the correlations between stripes in neighboring layers,\cite{tran96a} while the intensity is also influenced by the structure factor.  The inelastic scans at $E=3$~meV in Fig.~\ref{fg:Ldep25}b show that there is $L$-dependent structure in the inelastic scattering at temperatures near and above $T_{\rm co}$.  The modulated part of the signal we attribute to the structure factor associated with the fluctuating charge stripes.

\begin{figure}[t]
\centerline{\includegraphics[width=2in]{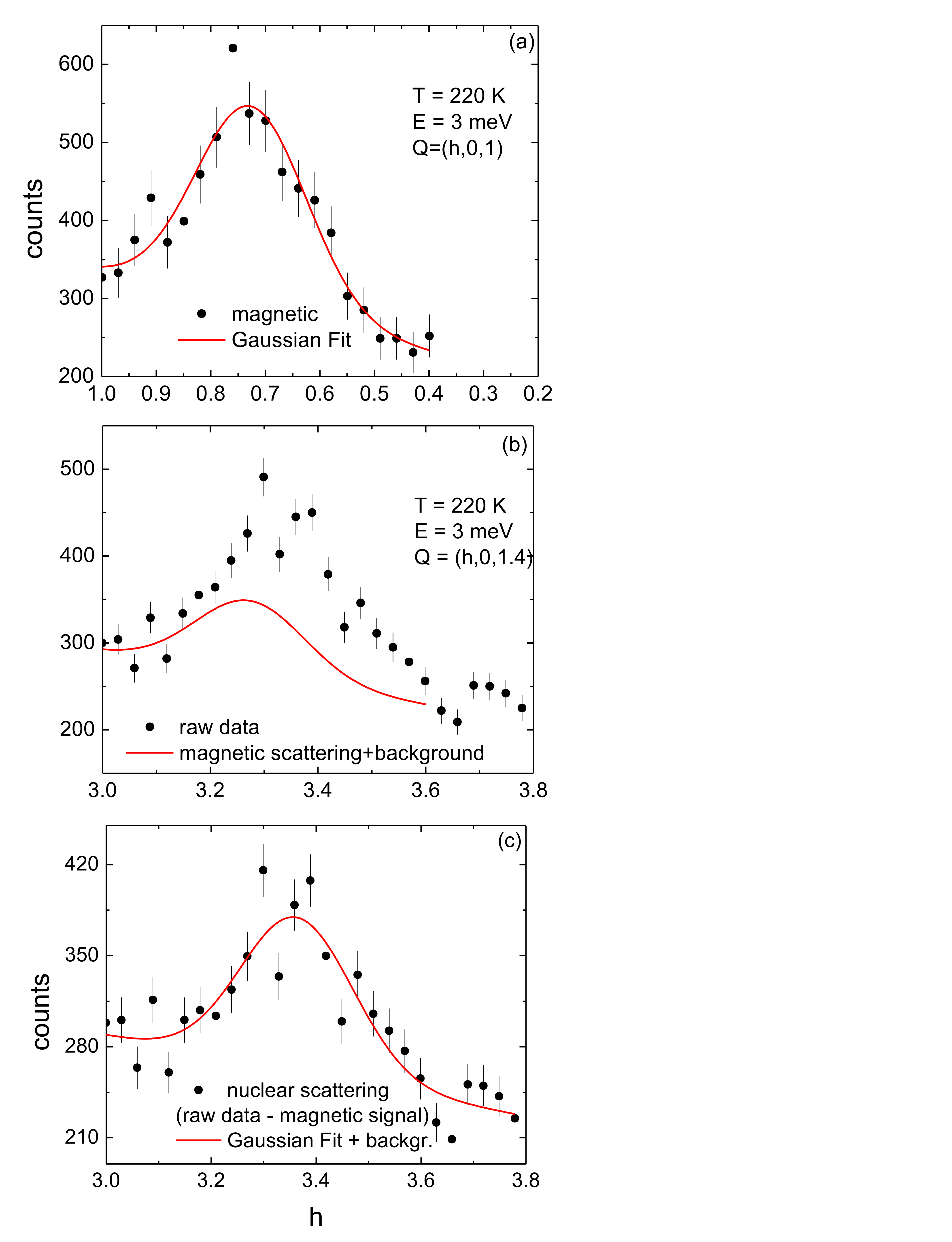}}
\caption{
{\fc  Spin and charge-stripe fluctuations at 220K for La$_{1.75}$Sr$_{0.25}$NiO$_4$.}  All scans are along ${\bf Q}=(h,0,1)$ at $E=3$~meV.
{\fc a} Magnetic scattering near $h=0.72$; red line is a fit of a Gaussian peak plus a linear background.  {\fc b} Inelastic scattering near $h=3.3$.  Red line is a scaled version of the Gaussian peak from (a), as described in the text.  {\fc c}  Data from (b) after subtracting the scaled magnetic contribution.  Red line is a fit of a Gaussian peak plus a linear background.
\label{fg:inel1}  }
\end{figure}

Because of broad $Q$ widths and shifts in $\delta$ with temperature,\cite{ishi04} it is still necessary to correct for magnetic scattering in order to establish the magnitude of the charge-stripe fluctuations.  The analysis process is illustrated in Fig.~\ref{fg:inel1}.  Again, we are analyzing the response at $E=3$~meV.  We take the magnetic scattering centered near ${\bf Q}_1'$ to be purely magnetic, and parametrize it as a Gaussian peak plus linear background.  We scale this contribution by $F^2=3.6$, obtained from the inelastic scans at 100~K shown in Fig.~\ref{fg:el1}c and d; the scaled result, plus background, is indicated by the solid curve in Fig.~\ref{fg:inel1}b.  Subtracting the scaled magnetic contribution yields the nuclear fluctuation scattering in Fig.~\ref{fg:inel1}c, from which we can determine the intensity of the charge-stripe fluctuations by fitting a Gaussian peak plus a linear background.  The temperature dependent data extracted in this way are plotted in Fig.~\ref{fg:Bragg_Inel}d.  

\bigskip\noindent{\fc Discussion}\par

Optical conductivity measurements\cite{kats96,home03,poir05,cosl13} of the stripe-ordered phase in La$_{2-x}$Sr$_x$NiO$_{4+y}$ reveal a broad peak centered at 0.6--0.8 eV, with a gap (measured at the peak half-height) about half that size, consistent with localization of the charge.   While nickelates develop a finite conductivity (qualifying them as bad metals) on warming through $T_{\rm co}$, a pseudogap remains, which only gradually decreases on heating to $\sim 2T_{\rm co}$.\cite{kats96,home03,cosl13,lloy08}  The pseudogap indicates a partial localization of charge at $T>T_{\rm co}$.  Indirect evidence for residual charge-stripe correlations was provided by a recent neutron powder diffraction study that found excess disorder of in-plane Ni and O sites whose temperature dependence was well correlated with the pseudogap.\cite{abey13}  Our results provide direct evidence that dynamic charge stripes survive above $T_{\rm co}$, and thus appear to resolve the nature of the pseudogap.  They also provide an explanation for the fact that spin fluctuations observed above $T_{\rm co}$ remain incommensurate\cite{tran97c,lee02,woo05}: the charge and spin stripes can maintain an instantaneous correlation even when static order is absent.  Similar conclusions regarding the coupling of spin and charge stripes and their survival as dynamic entities have been inferred from recent time-resolved pump-probe studies, involving an optical pump and a probe by resonant x-ray diffraction.\cite{lee12,kung13,chua13} 

In a terahertz spectroscopy study of La$_{1.67}$Sr$_{0.33}$NiO$_4$,\cite{lloy08} the temperature-dependent change in the conductivity at $E\approx 5$~meV was interpreted as evidence of impurity-pinned charge collective modes, by analogy with behavior in quasi-one-dimensional (quasi-1D) charge-density-wave (CDW) systems.\cite{grun88}  While such a picture would seem to be consistent with our observation of fluctuating charge stripes for $T\gtrsim T_{\rm co}$, we note that there are significant differences between charge stripes in a doped Mott insulator and CDW order in a quasi-1D metal.  In the former case, sliding of the charge stripes is prevented by the presence of the intervening spin stripes.  The fact that the magnetically-ordered Ni sites have $S=1$, rather than $S=\frac12$ in cuprates, helps to localize the charge stripes.\cite{zaan94,tran98c}  When the charge stripes start to move, the spin fluctuations become strongly damped.\cite{lee02,bour03,woo05}  In fact, the THz results\cite{lloy08} show that the low frequency conductivity begins a rapid growth near the spin-disordering temperature, $T_{\rm so}$, consistent with the rise in charge-stripe fluctuations found in our Fig.~\ref{fg:Bragg_Inel}(c).


The fact that the intensity of the fluctuations is strong for $T\gtrsim T_{\rm co}$ is consistent with the fluctuations representing a soft collective mode that turns into the charge-stripe order below $T_{\rm co}$.   It appears to be an example of an electronic nematic phase,\cite{kive98} which is correlated with a pseudogap.  The situation is rather different from a conventional Peierls transition in a quasi-one-dimensional metal.\cite{renk74}  

These low-energy excitations have the character of critical fluctuations, but with a broad critical regime on the disordered side of the transition.  We do not know of any calculations that properly describe the charge-stripe fluctuations that we observe.  Using a model appropriate for cuprates, Kaneshita {\it et al.}\cite{kane02} calculated charge collective modes that disperse steeply from the charge-order wave vector and leave detectable anomalies when they cross (and mix with) phonon branches, however there was no prediction of a low-energy lattice response independent of the phonon modes. Presumably, one must account for the strong electron-lattice coupling in nickelates.\cite{zaan94} The possible connection between the low-energy stripe fluctuations and the anomaly in a bond-stretching phonon branch\cite{tran02} and other possible phonon anomalies requires further exploration. 

Our results suggest that it may be possible to detect charge-stripe fluctuations in superconducting cuprates; however, this will be challenging because the smaller atomic displacements in the ordered state\cite{tran96b} suggest a weaker signal in the fluctuating state.  There have been recent optical pump-probe experiments on cuprates that have detected oscillations interpreted as evidence for collective excitations of charge-density waves\cite{torc13,hint13}; however, those measurements provide no information on the spatial periodicity of such modes. Inelastic neutron scattering has the potential to characterize both the spatial modulation and the dynamics of disordered charge stripes.

Finally, our observations have some relevance to the current discussion concerning CDW correlations\cite{wu11,ghir12,chan12a} in underdoped YBa$_2$Cu$_3$O$_{6+x}$.   Evidence for CDW order was initially detected by nuclear magnetic resonance (NMR) below the superconducting transition temperature, $T_c$, and at high magnetic field.  X-ray scattering measurements\cite{ghir12,chan12a} discovered broad CDW reflections in zero field, setting in at $T>2T_c$.  Under the assumption that all of the measurements are probing the same CDW correlations, it has been proposed that the x-ray measurements might be integrating over fluctuating CDW correlations, which would not be detectable by NMR.\cite{ghir12,chan12a}  There are problems with such an interpretation, however.  A recent study\cite{tham13} has shown that the integrated intensity of a CDW peak in YBa$_2$Cu$_3$O$_{6.6}$ is quite similar to one in charge-stripe-ordered La$_{1.875}$Ba$_{0.125}$CuO$_4$.  Could integrated fluctuations really provide as much intensity as static order?  In the case of La$_{1.67}$Sr$_{0.33}$NiO$_4$, our results directly demonstrate that charge-stripe fluctuations exist far above $T_{\rm co}$, whereas elastic x-ray scattering measurements\cite{du00} lost the signal by $T_{\rm co}+20$~K.  This strongly suggests that the x-ray studies of YBa$_2$Cu$_3$O$_{6+x}$ are measuring static correlations.  (Note that recent inelastic x-ray scattering measurements\cite{leta14} have narrowed the possible energy range of fluctuations to something less than 1~meV.)  As suggested in a recent theory paper,\cite{nie14} the short-range character of the CDW correlations seen with x-rays\cite{ghir12,chan12a} is likely a consequence of disorder.    NMR may have detected a transition to a distinct state, with longer-range order or a more commensurate wave vector.


\bigskip\noindent{\fc Methods}\par

\noindent{\bf Samples.} Single crystals of La$_{2-x}$Sr$_x$NiO$_4$ ($x = 0.33$, and also 0.25 and 0.45) were grown by the floating-zone method. 4~cm-long samples were cut from rods of 6~mm diameter. 

\bigskip
\noindent{\bf Neutron scattering experiments.}  Preliminary measurements as well as magnetic form factor measurement (Fig. 7) were carried out on the 1T spectrometer at the ORHPEE reactor at the Laboratoire Leon Brillouin at Saclay, France, using a fixed final energy of 14.8~meV. Inelastic neutron scattering experiments were carried out on the time-of-flight (TOF) Wide Angular-Range Chopper Spectrometer (ARCS) at the Spallation Neutron Source (SNS) at the Oak Ridge National Laboratory (ORNL) with the samples mounted with the $c$-axis vertical. Incident energies were 65meV with the chopper speed of 360Hz (data in figs. 2,3) and 60 meV with the chopper speed of 460Hz (data in Fig. 4). In order to map a region in the 4-dimensional reciprocal space, data were taken in the rotating sample mode. Measurements of the temperature dependence of low energy excitations (Fig. 1,5,6,8,9,10) were carried out using the HB-3 triple-axis spectrometer at the High Flux Isotope Reactor (HFIR), ORNL.



\bigskip

\noindent {\bf Acknowledgments}\ \    We are grateful for helpful comments from J. Zaanen.  S.A., D.P., and D.R. (G.D.G. and J.M.T.) were supported by the Office of Basic Energy Sciences, Division of Materials Sciences and Engineering, U.S. Department of Energy (DOE), through Contract No.\ DE-SC0006939 (DE-AC02-98CH10886).   The experiments at Oak Ridge National Laboratory's High Flux Isotope Reactor and Spallation Neutron Source were sponsored by the Division of Scientific User Facilities, US DOE Office of Basic Energy Sciences.

\noindent {\bf Competing Interests}\ \ The authors declare that they have no competing financial interests.

\noindent {\bf Author contributions}\ \ Sample preparation: G.D.G.; neutron scattering: S.A., D.P., K.M., M.D.L., S.C., J.F.B., D.L.A., D.L., J.M.T., D.R.;  paper writing: S.A., J.M.T., D.R.

\noindent {\bf Additional information}\ \  Correspondence and requests for materials
should be addressed to D.R.~(email: dmitry.reznik@colorado.edu).

\end{document}